\input harvmac
\input epsf
\noblackbox
\def\ias{\vbox{\sl\centerline{Theoretical Physics Group, Mail Stop 50A-5101}%
\centerline{Ernest Orlando Lawrence Berkeley National Laboratory}%
\centerline{Berkeley, CA 94720 USA}}}

\nref\one{G. Gibbons and P. Townsend, Phys. Rev. Lett. {\bf 71} (1993) 3754.}
\nref\two{J. Polchinski, Phys. Rev. Lett. {\bf 75} (1995) 4724.}
\nref\three{E. Witten, Nucl. Phys. {\bf B460} (1996) 335.}
\nref\four{A. Strominger and C. Vafa, Phys. Lett. {\bf B379} (1996) 99.}
\nref\five{I. Klebanov, Nucl. Phys. {\bf B496} (1997) 231.}
\nref\six{J. Maldacena, Adv. Theor. Math. Phys. {\bf 2} (1998) 231.}
\nref\seven{S. Gubser, I. Klebanov, and A. Polyakov; Phys. Lett. {\bf B428} 
(1998) 105.}
\nref\eight{E. Witten, Adv. Theor. Math. Phys. {\bf 2} (1998) 253.}
\nref\nine{I. Klebanov and A. Tseytlin, Nucl. Phys. {\bf B475} (1996) 179.}
\nref\ten{V. Balasubramanian and F. Larsen, Nucl. Phys. {\bf B478} (1996) 199.}
\nref\eleven{J. Maldacena, A. Strominger, and E. Witten; JHEP {\bf9712}
(1997) 002.}
\nref\twelve{J. Blum, unpublished.}
\nref\thirteen{E. Witten, Nucl. Phys. {\bf B202} (1982) 253.}
\nref\fourteen{J. Cardy, Nucl. Phys. {\bf B270} (1986) 186.}
\nref\fifteen{C. Callan and J. Maldacena,  Nucl. Phys. {\bf B472} (1996) 591.}
\nref\sixteen{J. Maldacena and L. Susskind, Nucl. Phys. {\bf B475} (1996) 679.}
\nref\seventeen{W. Taylor, Nucl. Phys. {\bf B508} (1997) 122.}
\nref\jdb{J. Blum, Nucl. Phys. {\bf B486} (1997) 34.}
\nref\eighteen{J. Schwarz, Nucl. Phys. Proc. Suppl. {\bf 55B} (1997) 1.}
\nref\nineteen{A. Sen, JHEP {\bf 9803} (1998) 005.}
\nref\twenty{M. Gaberdiel and B. Zwiebach, Nucl. Phys. {\bf B518} (1998) 151.}
\nref\twoone{O. Bergman, Nucl. Phys. {\bf B525} (1998) 104.}
\nref\sen{A. Sen, JHEP {\bf 9806} (1998) 007.}
\nref\twotwo{B. Simon, Ann. Phys. {\bf 146} (1983) 209.}
\nref\twothree{M. Claudson and M. Halpern, Nucl. Phys. {\bf B250} (1985) 689.}
\nref\twofour{U. Danielsson, G. Ferretti, and B. Sundborg; Int. J. Mod. 
Phys. {\bf A11} (1996) 5463.}
\nref\ms{S. Sethi and M. Stern, Phys. Lett. {\bf B398} (1997) 47.}
\nref\yi{P. Yi, Nucl. Phys. {\bf B505} (1997) 307.}
\nref\stern{S. Sethi and M. Stern, Commun. Math. Phys. {\bf 194} (1998) 675.}
\nref\hs{M. Halpern and C. Schwartz, Int. J. Mod. Phys. {\bf A13} (1998) 4367.}
\nref\pr{M. Porrati and A. Rozenberg, Nucl. Phys. {\bf B515} (1998) 184.}
\nref\mns{G. Moore, N. Nekrasov, and S. Shatashvili; hep-th/9803265.}
\nref\ko{A. Konechny, JHEP {\bf 9810} (1998) 018.}
\nref\wb{J. Wess and J. Bagger, {\it  Supersymmetry and Supergravity}
(Princeton University Press, 1992), chapter \uppercase\expandafter
{\romannumeral 7}, and citations there.} 
\nref\poly{A. Polychronakos, Phys. Lett. {\bf B408} (1997) 117.}
\nref\calo{F. Calogero, J. Math. Phys. {\bf 10} (1969) 2191 and 2197;
{\bf 12} (1971) 419.}
\nref\gibto{G. Gibbons and P. Townsend, Phys. Lett. {\bf B454} (1999) 187.}
\nref\spcalo{J. Gibbons and T. Hermsen, Physica {\bf D11} (1984) 337\semi
S. Wojciechowski, Phys. Lett. {\bf A111} (1985) 101.}
\nref\adsone{P. Claus, M. Derix, R. Kallosh, J. Kumar, P. Townsend, and
A. Van Proeyen; Phys. Rev. Lett. {\bf 81} (1998) 455.}
\nref\polym{J. Minahan and A. Polychronakos, Phys. Lett. {\bf B326} (1994) 
288.}
\nref\adsbps{J. Michelson and M. Spradlin, hep-th/9906056\semi
S. Corley, hep-th/9906102\semi J. Lee and S. Lee, hep-th/9906105.}
\nref\adsact{J.-G. Zhou, hep-th/9906013.}

\Title{\vbox{\baselineskip12pt
\hbox{LBNL-43734}\hbox{hep-th/9907101}}}
{\vbox{\centerline{Supersymmetric Quantum Mechanical}\centerline
{Description 
of Four Dimensional Black Holes}}}

{\bigskip
\centerline{Julie D. Blum}
\bigskip
\ias

\bigskip
\medskip
\centerline{\bf Abstract}
By assuming the existence of a novel multipronged string state for 
D-particles interacting with D-brane intersections in type IIA string
theory, we are able to derive a quantum mechanical description of
supersymmetric Reissner-Nordstrom black holes.  A supersymmetric
index calculation provides evidence for this conjecture.  The 
quantum mechanical system becomes two decoupled conformal quantum mechanical 
systems in the low energy limit.  The conformal quantum mechanics has expected 
properties of a dual description of string theory on $AdS_2\times S^2$.
}

\Date{6/99}

\newsec{Introduction}

Following work describing the near horizon geometry of certain string (M)
theory black holes composed of solitonic branes as the maximally supersymmetric
product of Anti de Sitter space and a sphere
($AdS\times S$) with a conformal theory on the boundary \one , work
that identified certain of the solitonic black holes as Dirichlet(D)-branes
\two , work
determining the low energy theory of D-branes to be a nonabelian gauge theory
\three\ known to be conformal in certain cases, and work showing that
calculations of the properties of D-brane black holes could be performed
successfully in the conformal theory for an appropriately large number of
D-branes \four\five ; there was a conjecture \six\ that supergravity or
string (M) theory in the near horizon $AdS$ geometry of the solitonic (D-)
branes was equivalent to the conformal theory on these branes.  Further
work gave a recipe for comparing the two theories and provided some
evidence for the conjecture's validity in the supergravity limit \seven\eight .
Whether or not all of the interesting aspects of string theory can be reduced
to a field theory, one can at least derive some useful relations between the 
two theories following the above works.

The aim of this paper is to extend the relation to four-dimensional
black holes with a near horizon geometry of $AdS_2\times S^2$.  We will
show that the two-dimensional conformal theory descriptions of the
onebrane-fivebrane black hole \four\ and generalizations \nine\ten\eleven\
are alternatively described at low energies by a quantum mechanics that 
becomes conformal in the very low energy limit.  Evidence will be presented 
that 
this quantum mechanics contains the degrees of freedom responsible for the 
ground state entropy of the black holes.  The quantum mechanics 
will not describe completely the 
moduli space of the transverse six-fold, for we will assume that the
local geometry of the D-particle 
is flat, and we will also neglect background fields.  A 
two-dimensional description \twelve\four\eleven\ may be better 
suited for this purpose although one could further complicate the quantum 
mechanics.  On the other hand, to understand macroscopic features of the 
four-dimensional black hole, this quantum mechanics may be a reasonable 
approach.  In the course of obtaining the quantum mechanical theory, we
will propose some novel string states occurring at the intersections 
of D-branes.  We hope that this proposal leads to a better understanding of 
these intersections.  The rules we will develop are somewhat {\it ad hoc}
but seem to lead to a sensible description.

The outline of the rest of this paper is as follows.  In section two we
will review some macroscopic properties as well as the microscopic effective
string formulation of the black holes to be discussed.  In section three
we will present the novel string states that we believe to capture the low
energy degrees of freedom of the black holes and a prescription for 
obtaining these states from the intersections of D-branes.  In section
four we will calculate the index of supersymmetric ground states \thirteen\
of the quantum mechanics in the simplest theory containing these states.
We will extrapolate from this result a conjecture for the degeneracy of
the large number of intersections case.  The resulting ground state
entropy will agree with the macroscopic and string formulation
predictions.  In section five we will derive the quantum mechanical system
describing the black holes.  We will take the low energy limit and
obtain a conformal quantum mechanics.  What is interesting here is that
in this limit we appear to have two decoupled  conformal
quantum mechanical systems, a ``Coulomb'' branch with manifest $SO(3)$
symmetry and a ``Higgs'' branch with a large internal symmetry.  However, 
these two 
branches are coupled in the full nonconformal theory.  
In section six we present our
conclusions and directions for further research.

\newsec{Black Holes and Effective Strings}

\subsec{Review of Macroscopic Black Holes}

The four dimensional black holes we will consider in this paper are
all extremal and of the Reissner-Nordstrom type.  The metric takes the
following form:
\eqn\rn{ds^2=-{(T_1 T_2 T_3 T_4)}^{1/2}dt^2 +{(T_1 T_2 T_3 T_4)}^{-1/2}
(dr^2 +r^2 d{\Omega}_2 ^2)}
where $T_i =(1+Q_i /r)^{-1}$ and the $Q_i$ are positive.
There are electric and magnetic fields,
\eqn\fie{F_i=dt\wedge dT_i + \ast dT_i^{-1}.}
The mass of the black hole is 
\eqn\mass{M={\sum_i Q_i\over 4 G_N}}
with $G_N$ the four-dimensional Newton constant.  For equal charges
$Q$ the Ricci scalar vanishes and
\eqn\ric{R_{\mu\nu}R^{\mu\nu}={4Q^4\over (r+Q)^4}}
so there is no singularity in the extremal limit.  The extremal
entropy is
\eqn\s{S={\pi{(Q_1 Q_2 Q_3 Q_4)}^{1/2}\over G_N} .}
In the near horizon Reissner-Nordstrom reduces to $AdS_2\times S^2$
with metric
\eqn\rnnh{ds^2_{NH}={-r^2\over {(Q_1 Q_2 Q_3 Q_4)}^{1/2}}dt^2 +
{{(Q_1 Q_2 Q_3 Q_4)}^{1/2}\over r^2} dr^2 +
{(Q_1 Q_2 Q_3 Q_4)}^{1/2}d{\Omega}_2 ^2}
while the metric at infinity is flat.
The Ricci scalar of $AdS_2$ is $R={-2\over {(Q_1 Q_2 Q_3 Q_4)}^{1/2}}$,
and the cosmological constant is $\Lambda ={1\over 2}R$ while
$R_{S^2}=2\Lambda_{S^2}=-R_{AdS_2}$.  There are numerous papers
that have studied the Reissner-Nordstrom metric as a solution of 
string (M) theory.  In type IIB string theory in a purely threebrane
background, the equation to solve is 
\eqn\ffive{R_{\mu\nu}=F_{\mu a_1 a_2 a_3 a_4}F_{\nu}^{\,\, a_1 a_2 
a_3 a_4}}
where we have distinguished four-dimensional and six-fold indices in an 
obvious way, and $F$ is the five-form field strength.  In the simplest
case (a six-torus), one can reverse the signs of some components of
the field strength while retaining a solution of the low energy field
theory.  Some of these reversals will break supersymmetry, and it is
interesting to consider these black holes.  We will comment on the 
consequences of breaking supersymmetry in this way in the next section.

\subsec{D-branes and Microscopic Strings}

The paradigmatic extremal onebrane-fivebrane black hole \four\ is
five-dimensional with near horizon geometry $AdS_2\times S^3$.  Upon
compactification on a circle the geometry is again Reissner-Nordstrom.
This black hole has three charges.  In the D-brane approach, these charges
are the number $N_5$ of fivebranes wrapped on $K3\times S^1$, the number
$N_1$ of onebranes wrapped on $S^1$, and the momentum $p=N_0 /R$ with $R$
the radius of the $S^1$ and $N_0$ an integer.  At low energies there is
an effective conformal theory on $S^1\times time$ with central charge 
$c=6N_1 N_5$.  The momentum $N_0$ corresponds to the eigenvalue of the
Virasoro generator $L_0$.  The sign of $N_0$ is not crucial here as the
theory is left-right symmetric, but the signs of $N_1$ and $N_5$ are
correlated.  The entropy has been calculated \four\ in the limit of large
charges using the metric of the supergravity solution and alternatively 
the asymptotic
microscopic formula for the degeneracy \fourteen , 
\eqn\deg{d(N_0 ,c)=exp{2\pi\sqrt{{N_0 c\over 6}}}}
where $d$ is the degeneracy, and the entropy 
$S=ln d=2\pi\sqrt{{N_0 c\over 6}}$.  The two calculations of the entropy 
agree.  The $K3$ can be replaced by a $T^4$
\fifteen\sixteen\ with similar results.  For the $T^4$ case one can
choose any combination of signs for the three charges.

By a sequence of U-duality operations we can convert the $T^4$ case to
an M theory configuration.  Compactify on a circle to four dimensions.
Perform a T-duality on three directions--the newly compactified direction
and two directions of $T^4$ (avoiding the momentum circle).  
Interchange the M theory circle and the momentum circle.  The result is $N_1 
N_5$ fourbrane intersections on a two-torus and $N_0$ zerobranes.  We 
presumably can play the same game with $K3$ using mirror symmetry but the
analysis seems more complicated for this case.  In the latter part of
this paper we will derive an effective quantum mechanics 
for the D-particles at the intersections.

The other prototypical Reissner-Nordstrom black hole solution of string
theory has been discussed by \nine\ten\ and many others.  In type IIB
string theory the four charges $N_i$ are due to threebranes wrapped
on a $T^6$ so that any two sets $(i,j)$ intersect in a string while
three or four sets intersect in a point.  There are therefore six strings
along each direction $(T_{(i)}^3\cap T_{(j)}^3)$ of the six-torus and a
total of $N_1 N_2 N_3 N_4$ intersections on $T_{(1)}^3\cap T_{(2)}^3\cap
T_{(3)}^3\cap T_{(4)}^3$.  The supersymmetries which are preserved
satisfy the following conditions:
\eqn\super{\eqalign{& \Gamma_{11}\epsilon_L =\epsilon_L\cr
& \Gamma_{11}\epsilon_R =\epsilon_R\cr
& \Gamma_{0abc}^i\epsilon =\pm i\epsilon\cr}}
where $\Gamma_{11}=\Gamma_0 \Gamma_1\dots\Gamma_{10}$ with $\Gamma_a$
a ten-dimensional Clifford algebra matrix, $\Gamma_{0abc}^i=\Gamma_0 \Gamma_a
\Gamma_b \Gamma_c$ where $a,b,c$ are the directions on $T^6$ on which the
$N_i$ branes are wrapped,  
$\epsilon_L$ and $\epsilon_R$
are the two supersymmetries of type IIB from left and right movers of
the string, and $\epsilon=\epsilon_L +i\epsilon_R$.  
The sign of the last relation depends on the sign of $N_i$.  The 
$\Gamma^i$ commute and satisfy $\Gamma^1 \Gamma^2 \Gamma^3 \Gamma^4=\pm 1$,
and the number of preserved supersymmetries is
\eqn\supergood{N=tr[{1\pm i\Gamma^1\over 2} {1\pm i\Gamma^2\over 2}
{1\pm i\Gamma^3\over 2} {1\pm i\Gamma^4\over 2}].}
Thus, $N=4$ or $N=0$.  Regardless of the signs, any triple intersection
preserves supersymmetry, and supersymmetry can be broken only on
quadruple intersections.  Since the nonsupersymmetric case solves the
low energy equations of IIB with Reissner-Nordstrom geometry, one might hope
to find a conformal quantum mechanical dual for the near horizon geometry.
Unfortunately, an analysis \seventeen\ reveals that the nonsupersymmetric
configuration does not minimize the energy and is expected to be unstable.

By T-dualizing this configuration we obtain $N_2 N_3 N_4$ intersections of
M theory fivebranes on an effective string with $N_1$ units of 
momentum.  With the proper normalization of charges, the entropy has
been calculated macroscopically \s\ to be $S=2\pi\sqrt{N_1 N_2 N_3 N_4}$,
and arguments have been given that this result holds microscopically
\nine\ten .  By deforming the degenerate fivebranes into a smooth fivebrane,
one can use the prescription of \eleven\ to determine the microscopic 
entropy.  One finds that $c_L=c_R=6N_2 N_3 N_4$ up to a negligible
correction for large charges, and the entropy agrees with the macroscopic 
prediction.  The left side is almost entirely bosonic, whereas the
right side is supersymmetric.  Because of the asymmetry of left and
right movers, we can view the nonsupersymmetric instability associated
with the wrong sign momentum as a tachyon.  From the type IIA perspective
the momentum is equivalent to $N_1$ D-particles whose quantum mechanics
we will derive.  We expect that this quantum mechanics will apply to
any supersymmetric black hole with a Reissner-Nordstrom
metric when one ignores corrections based on the transverse 
six-fold geometry.   The full geometry possibly can be incorporated in this
quantum mechanics, but the analysis is not within the scope of this
paper.

\newsec{Multistrings at D-Brane Intersections}

In this section we conjecture that the states describing D-particle 
interactions at the intersections of D-branes are multipronged fundamental 
strings
that attach to the D-particle.  In string theory novel states
are sometimes required in special compactifications such as the twisted open
strings discovered \jdb\ in the context of certain orientifolds.  
The idea of multipronged 
strings (multistrings)
previously found an application in type IIB string theory to describe
certain BPS states \eighteen\nineteen\ including states responsible
for exceptional gauge symmetries \twenty\ and nonperturbative states
preserving one-quarter of the supersymmetry in $N=4$ Yang-Mills 
theory \twoone .  The context in which we are proposing these objects
is novel.  The considerations of this section are the most conjectural
of this paper as we will not at this time try to prove the existence
of these objects.  Our main argument for invoking these states is that
they lead to a quantum mechanical description that satisfies many requirements
of a dual to supersymmetric string theory in the background of a
Reissner-Nordstrom black hole.  Another argument is that we expect the 
low energy degrees of freedom to carry the charges of all the branes 
at the intersection.  A perhaps more prosaic consideration is 
that an index theory calculation similar to that of section four was attempted
by assuming the presence of the usual D0-D4 matter at these intersections.
Not only was the calculation formidably impossible (for me) but also an
upper bound on the integral seemed to be too low.  By contrast, the 
calculation with these states is a piece of cake and yields the desired 
result.  The natural assumption is that multistrings are the
bound (BPS) states of the intersection.

Let us now describe these states.  We assume that $n$ D-branes of the same 
dimension $d$ intersect along some locus such that any two D-branes
are orthogonal to each other (can at most intersect in less than $d$
dimensions) and some supersymmetry is preserved.  The BPS state that
we conjecture has $n+1$ prongs, one end on a D-particle and one end
on each of the $n$ branes.  Our assumption is that there is always 
such a string with an endpoint carrying charge $|q|=1$ under the 
$U(1)$ gauge group
of the D-particle.  By symmetry each of the $n$ D-branes must
contribute $|q|=1/n$ to this charge.  Such a string will have at 
least three prongs (the case $n=1$ is the usual case) and break the
supersymmetry from $32$ to no more than $4$ supercharges.  We then
assume that other states for $n\ge 3$ can be obtained by reversing 
the orientations of an even number of the $n$ strings attached to the 
$n$ different branes. 

We show these states for $n=2$ and $n=3$ in figure one.  If the string 
entering the D-particle carries no charge, the state can 
be deformed to one that does not interact with the D-particle and should
not be considered in the quantum mechanics.  Although we have drawn the
strings with finite size for clarity, these states are massless and can
shrink to a point as appropriate for low energy modes of the quantum
mechanics.  We assume that the rule requiring the number of orientation 
changes to be even is related to the fact \supergood\ that supersymmetry 
is broken for an odd number of brane orientation changes for $n\ge 3$.
Here we are fixing the D-particle orientation.
\vfill\eject

%Figure 1
\centerline{\epsfxsize=1\hsize\epsfbox{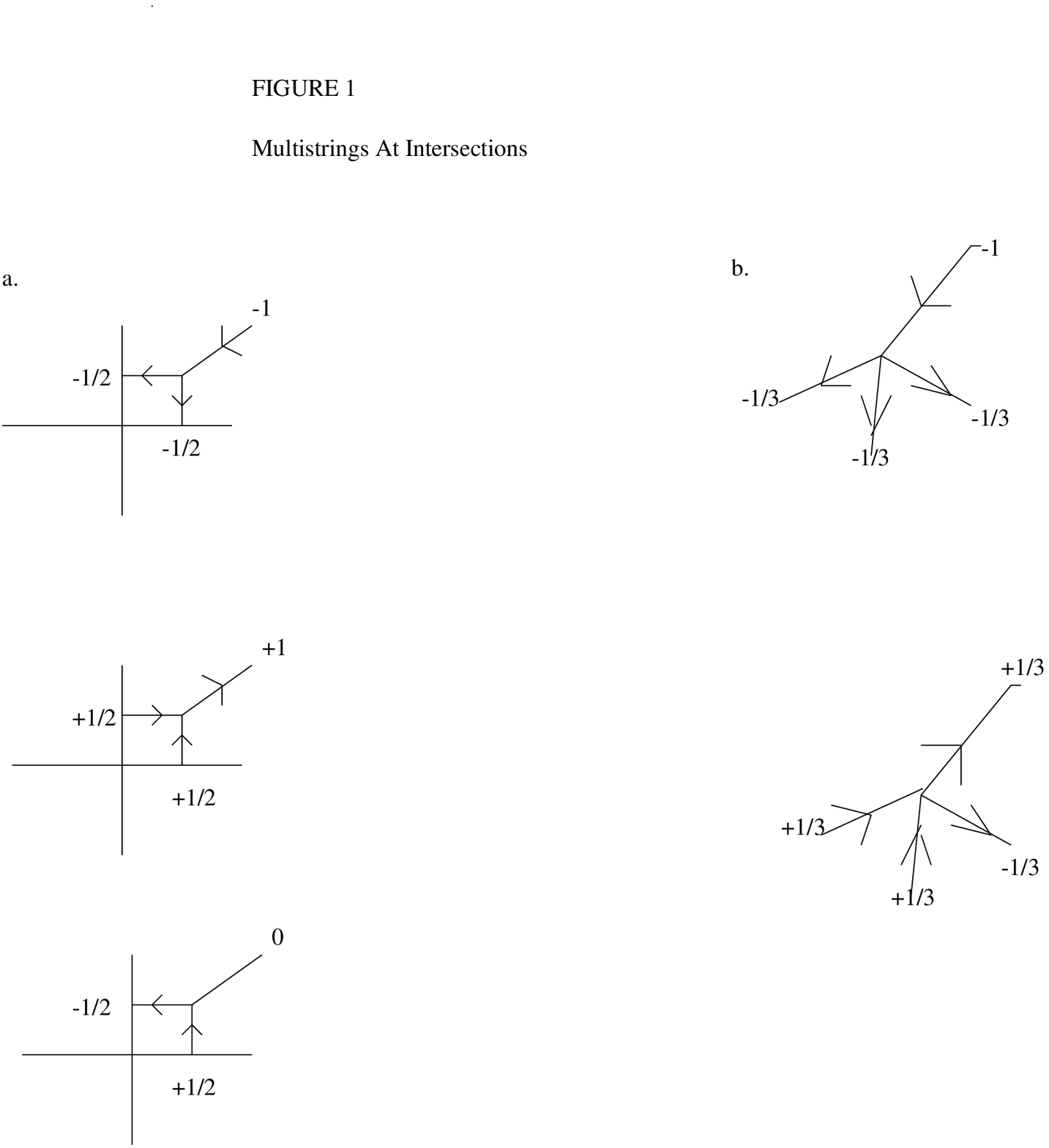}}
\bigskip
\centerline{\vbox{\noindent{\bf Fig. 1.}
a. Multistrings localized at an $n=2$ intersection.
b. Multistrings at an $n=3$ intersection.  There are $3$ multistrings
with charge $+1/3$.}}
\vskip .7cm

When the intersection of pairs of D-branes has dimension greater than zero,
we assume that there is a multistring at the intersection with ends
on each of these intersections such that $|q|=2$.  Again we can change
an even number of the ${n!\over (n-2)! 2!}$ orientations to obtain
other states as shown in figure two for $n=3$.  We can iterate this process
to higher intersections which are possibly significant for 
compactifications to less than four dimensions, but in four dimensions
the process ends with pairs.  Our main consideration here is that the 
minimal set of objects required by symmetry between the branes is invoked.
There are two overall signs both here for the charge and in section four
for the index.  The overall sign for the charge will not affect the 
calculation.  We are assuming a specific choice of overall sign for the
index
that yields a result consistent with expectations for the supersymmetric
Reissner-Nordstrom black hole.

\vskip .6cm
%Figure 2
\centerline{\epsfxsize=1\hsize\epsfbox{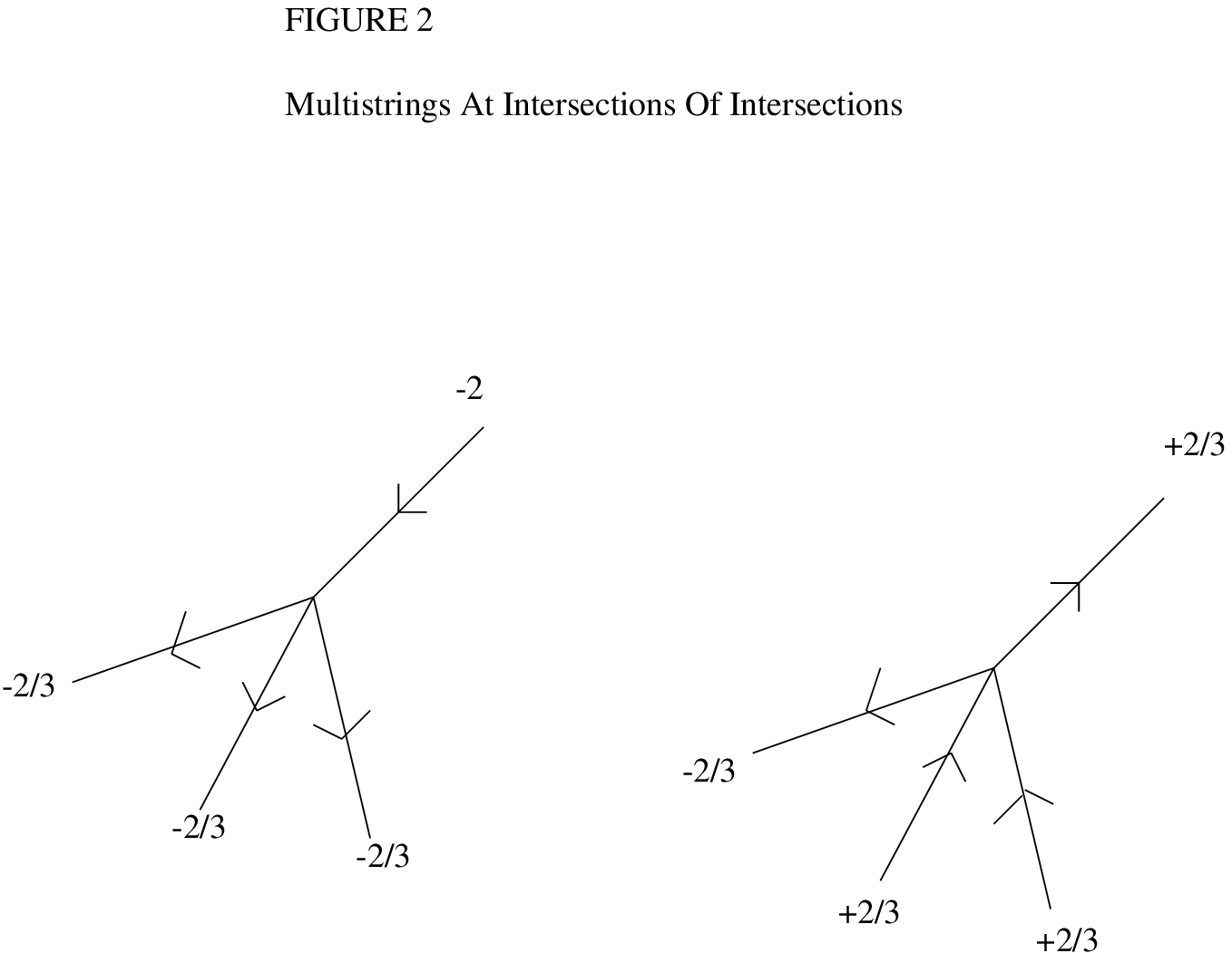}}
\bigskip
\centerline{\vbox{\noindent{\bf Fig. 2.}
There are $3$ $+2/3$ charged multistrings.}}
\vskip .7cm

Let us further argue for the plausibility of the multistring states.
If one considers two fourbranes intersecting in a membrane, the gauge
theory on the membrane is $U(1)\times U(1)$ with hypermultiplets
charged concurrently under the two $U(1)$'s.  This theory is the same
as that for a membrane transverse to a ${\bf R}^4/{\bf Z}_2$ orbifold in
type IIA.  This observation suggests an analogy between intersections
of D-branes and orbifolds.  If one thinks about the intersection of
two fourbranes as a degenerate limit of one smooth fourbrane, the
curvature of this fourbrane is not well-defined at the intersection locus.
This locus is analogous to the collapsed two-cycle at the ${\bf Z}_2$
fixed point of the orbifold.  At a ${\bf Z}_2$ fixed point one can
obtain half-integer values for the Neveu-Schwarz antisymmetric two-form.
Branes which lack moduli to move away from the fixed point can be
interpreted as branes with two extra dimensions wrapped around the
collapsed two-cycle at the fixed point.  The ``dimensional'' reduction
of couplings of the gauge field strength and two-form in the two
``extra dimensions'' can induce a half-integral charge for the endpoint
of a fundamental string on a brane that is stuck at the fixed point.  
An endpoint
that can move away must be integrally charged (the image brane must 
exist at the fixed point).  If a string endpoint does carry half-integral
charge, there must be another endpoint stuck at the fixed point
carrying half-integral charge by charge conservation.

Interpreting the intersection locus as a fixed locus of an orbifold, we
are freezing all moduli that move D-branes of spatial dimension greater
than zero away from this locus.  (One can also make an analogy between
string worldsheet orbifolds (orientifolds) and D-brane worldvolume
orbifolds.)  Before adding the D-particles there are no fractional
charges, and one has the usual theory of the intersection.  The D-particles
add magnetic flux (the D-particle is an instanton in each of the 
intersecting branes) to the intersecting fourbranes which is localized
at their position.  At the intersection there is a ${\bf Z}_2$
symmetry, exchanging the intersecting branes.  The orbifold analogy
suggests that the instanton flux due to the D-particle can take 
half-integral values in each one of the intersecting membranes
at the intersection so that the endpoints of the multistring at the 
intersection carry half-integral charge.  The Ramond-Ramond gauge
field on the D-particle couples to the total instanton flux from
the intersecting branes.  Since the D-particle is a point, the other
endpoint of the multistring must provide a cancelling flux.
This argument is sensible when all the endpoints are at the intersection.
We might expect that massive charged states in the D-particle quantum
mechanics do not respect supersymmetry.  The analysis of the ``Coulomb''
branch in section four confirms this expectation.  The D-particles
must couple to uncharged combinations of multistrings in leaving the
intersection.

One can generalize the above remarks to the case of three fourbranes
intersecting in a point where the intersection is invariant under
${\bf S}_3$ permutations of the intersecting branes.  Because of this
symmetry, it is plausible that the instanton flux of the intersection
and therefore the charge can be quantized in units of $1/3$.  The
main point is that intersections as a singular limit of smooth D-branes
should contain possibly extra massless states localized at the singularity just
like what has been found for the other singularities of string theory. 

We now come to a crucial distinction between the $n=2$ and $n=3$ cases.  
A supersymmetric massive deformation of the $n=2$ multistrings along
the compact directions is possible since there are two multistrings of
opposite charge.  The orbifold analogy suggests that the ${\bf Z}_2$
symmetry of the intersection should be preserved in this deformation.
Note that in this deformation only the endpoint attached to the D-particle 
leaves the intersection.  (Figure 1a is a little misleading.)
The $n=2$ case corresponds
to two fourbranes intersecting along a two-torus.  Assume that brane one
is wrapped on $T_{(1)}^2\times T_{(2)}^2$ while brane two is wrapped on
$T_{(1)}^2\times T_{(3)}^2$.  Let the intersection have complex coordinates
$y_2=y_3=0$.  There is a BPS deformation direction obtained by
requiring $|y_2|=|y_3|$.   The mass will be determined
by $|y_2|$ so there is an extra $S^1$ that decouples in addition
to $y_1$.  This $S^1$ degenerates at the intersection which seems to pose
a problem for the counting of states.  We will show in the next section
that this apparent problem does not exist.  Note that having obtained
this result, the BPS spectrum of the D-particle for the $U(1)$ case
is almost identical to that for the D0/D4 bound state problem.  The
counting of states in the next section will be facilitated by
this observation.

Our ``rules'' give nice results for the counting of states and seem to
be logical, but we cannot rule out a different set of states giving
equally nice results and being the correct states.  If this 
turns out to be the case, we are consoled
by the fact that the ``Coulomb'' branch of the conformal quantum
mechanics (to be derived in section five) should be unchanged.  It
will be interesting to see whether one can put the existence of these 
multipronged IIA strings at brane intersections on a firmer footing
(perhaps by relating them to M theory membranes ending on fivebranes).

\newsec{Bound States at Threshhold and Counting of Microscopic BPS States}

In this section we will calculate the index of supersymmetric 
ground states in the simplest versions of the theories we have postulated
in the previous section.  Our result will provide evidence for the formulas 
we will conjecture for the general case.  The calculation will involve 
bound states at threshhold, and some of the previous relevant work
includes \thirteen , \refs{\twotwo - \ko}.  Our calculations will be 
similar to the ones given in \ms\yi\stern .

\subsec{Setting up the Calculation}

We will study the case of one D-particle interacting with one intersection
of fourbranes having $n=2$ or $n=3$.  By our proposal of section three,
this theory is a quantum mechanics with four supercharges which can be
obtained from the dimensional reduction of the $N=1$ Yang-Mills theory
in four dimensions.  The formulas of \wb\ are particularly useful in this 
regard although we will make some changes in their conventions following \ms .
Let us first deal with the $n=2$ case.  This theory is a $U(1)$
gauge theory with two chiral multiplets having charge $q=\pm 1$.  There is
also an uncharged chiral multiplet that interacts with the charged multiplets 
via a superpotential.  Additionally, there are some decoupled degrees of 
freedom.  In calculating the index the gauge coupling constant $e$ can be
set to any nonzero value as it scales out of the index calculation. 
For the purpose of this computation we will
set $e=2$ in the Lagrangian of \wb .  The Lagrangian
also depends on another coupling constant $g$ for the superpotential
term.  Unlike the case of \ms\ we are considering the dimensional 
reduction of an $N=1$ not $N=2$ theory so the value of $g$ is not set
by supersymmetry.  Nevertheless, the calculation of the index cannot
depend on this value so long as it is nonzero, and the calculation is
simplest when we choose $g=\sqrt{2}$ as in the $N=2$ case.  We have 
argued in section three that $g$ should be nonzero.  With these choices
the Hamiltonian takes the following form after replacing the nondynamical
variables $D$ and $F$ \wb\ by the values that solve their equations of motion.
\eqn\ham{\eqalign{H=&{1\over 2} p^i p^i+p_y p_y^{\dagger}+p_+ p_+^{\dagger}
+p_- p_-^{\dagger}
+{1\over 2}{(Q_+ Q_+^{\dagger}+Q_- Q_-^{\dagger})}^2\cr 
&+(x^i x^i + 2yy^{\dagger})(Q_+ Q_+^{\dagger}+Q_- Q_-^{\dagger})+H_F\cr }}
\eqn\hamf{\eqalign{H_F&=x^i (M_+^{\dagger}\sigma_i M_+ - M_-^{\dagger}
\sigma_i M_-)
+\sqrt{2}(y M_- uM_+ -y^{\dagger}M_-^{\dagger} u M_+^{\dagger})\cr
&
+\sqrt{2}(Q_+ M_+^{\dagger}uL^{\dagger}-Q_+^{\dagger} M_+ uL-
Q_- M_-^{\dagger}uL^{\dagger}+Q_-^{\dagger} M_- uL)\cr
&
+\sqrt{2}(Q_- M_+ uN +Q_+ M_- uN
-Q_+^{\dagger} M_-^{\dagger}uN^{\dagger}
-Q_-^{\dagger} M_+^{\dagger}uN^{\dagger})\cr}}
\eqn\cons{iC_B=Q_-^{\dagger}p_-^{\dagger}-Q_+^{\dagger}p_+^{\dagger}
+Q_+ p_+ -Q_- p_-}
\eqn\consf{C_F=M_+^{\dagger}M_+ - M_-^{\dagger}M_-}
where the momenta are $p^i={\delta {\cal L}\over \delta {\dot x^i}}$, etc. ,
$x^i$ are the spatial components in the reduction of the four-dimensional
gauge fields, $\sigma_i$ are the usual Pauli matrices, $u=-i\sigma_2$,
$y$ is the complex scalar in the neutral chiral multiplet, $Q_{\pm}$
are the complex scalars in the charged chiral multiplets with charges
$q=\pm 1$, and $C_B$, $C_F$ are the bosonic and fermionic constraints 
generating gauge rotations (${\delta {\cal L}\over  \delta A_0}=C_B+C_F$).
We have used $\dagger$ as hermitian conjugate or complex conjugate
depending on the context, and we have chosen the gauge $A_0 =0$.
The complex two-component fermions $L$, $M_+$, $M_-$, $N$ satisfy the
anticommutation relations,
\eqn\anticom{\{ L_{\alpha}, L_{\beta}^{\dagger} \} =\delta_{\alpha\beta}.}

Next let us consider the $n=3$ case.  This theory is a $U(1)$ gauge theory
with three chiral multiplets of charge $+{1\over 3}$, three with charge
$+{2\over 3}$, one with charge $-1$, and one having charge $-2$.  Note
that as a four-dimensional theory there would be an anomaly, but this
anomaly is irrelevant for the quantum mechanics.
{\it A priori} we have the possibility of a superpotential coupling
together three chiral multiplets of $U(1)$ charges ${1\over 3}$, ${2\over 3}$,
and $-1$.  This superpotential could lift some or all of the flat
directions of the ``Higgs'' branch.  We will assume here that the 
superpotential is absent.  One reason is that the coupling together
of these charges presumably can be deformed into an object not localized at
the intersection.  Another reason is that the index calculation becomes
extremely difficult with a superpotential. Actually, we will have a more
concrete statement about
a superpotential when we discuss the index calculation.
Once we have turned off the
superpotential, we are guaranteed by supersymmetry in four dimensions
that there will be no perturbative or nonperturbative (for the U(1)
case) corrections.  In the quantum mechanics holomorphy should also
ensure that this coupling remains zero. 
The Hamiltonian and constraints for the $n=3$ case then are
\eqn\hamthree{\eqalign{H=&{1\over 2} p^i p^i+\sum_m p_m p_m^{\dagger}
+{1\over 2}{(\sum_m q_m Q_m Q_m^{\dagger})}^2 \cr
& +x^i x^i\sum_m q_m^2 Q_m Q_m^{\dagger}+H_F \cr}}
\eqn\hamthreef{H_F = \sum_m q_m M_m^{\dagger}\sigma\cdot x M_m
 +\sqrt{2}(\sum_m q_m Q_m M_m^{\dagger}uL^{\dagger}+h.c.)}
\eqn\conthreeb{iC_B=\sum_m q_m Q_m p_m -h.c.}
\eqn\conthreef{C_F=\sum_m q_m M_m^{\dagger}M_m}
where $m$ indexes the chiral multiplets, $q_m$ is the charge, and $h.c.$ is
the hermitian conjugate.  We can write the supersymmetries as
\eqn\supers{\eqalign{Q_{\alpha}=&{(\sigma\cdot p L)}_{\alpha}-\sqrt{2}\sum_m 
{(uM_m^{\dagger})}_{\alpha}p_m^{\dagger}\cr
&+ \sqrt{2} i\sum_m{(\sigma\cdot x uM_m^{\dagger})}_{\alpha}Q_m q_m
-iL_{\alpha}\sum_m Q_m Q_m^{\dagger}q_m\cr}}
where
\eqn\supercom{\{ Q_{\alpha}^{\dagger},Q_{\beta}\} =2\delta_{\alpha\beta}H
-2{(\sigma\cdot x)}_{\alpha\beta}(C_B+C_F)}
and
\eqn\supercomtwo{\{ Q_{\alpha},Q_{\beta}\} =0.}

We now outline the index calculation.  Following \thirteen\ the goal is 
to calculate the supersymmetric index or partition function with
the insertion of $(-1)^F$,
\eqn\index{I=\lim_{\beta\rightarrow\infty}I(\beta )=
\lim_{\beta\rightarrow\infty}Tr (-1)^F e^{-\beta H}=n_B -n_F .}
This index computes the number of bosons minus the number of fermions
in the supersymmetric ground state where $F$ is fermion number.  
The sum is only over gauge invariant states.
If the spectrum were discrete, there would be no dependence on $\beta$.
When there is a continuous spectrum above the ground state, the 
density of bosonic and fermionic states can differ and depend on $\beta$.
The usual procedure is to calculate the index as a sum of two terms,
$I=I(0)+\Delta I$ where
\eqn\deli{\Delta I=\int_0^{\infty}{d\over d\beta}I(\beta).}
The partition function can be constructed perturbatively in powers of 
$\beta$ so that the $\beta\rightarrow 0$ limit is easily taken.  The 
boundary correction $\Delta I$ is more subtle.  The bosonic potential
has noncompact (flat) directions along which this potential vanishes.
Near these directions the hamiltonian is a supersymmetric harmonic
oscillator in the transverse directions.  The frequency of the oscillator
increases linearly with distance from the origin along a flat direction,
but the ground state energy of the supersymmetric oscillator vanishes.
One can therefore have finite energy scattering states along these directions
so that the index depends on $\beta$, and there is a possible correction
$\Delta I$.

One includes a projection onto gauge invariant states $\int_{U(1)}d\theta
e^{i\theta C}$ where $C=C_B +C_F$ so that $I(\beta)$ becomes \ms\
\eqn\indexfor{I(\beta)=\int_{U(1)}d\theta \int dx \langle x|Tr e^{i\theta C} 
(-1)^F
e^{-\beta H}|x\rangle =\int_{U(1)}d\theta \int dx \langle gx|Tr \Pi (g)(-1)^F
e^{-\beta H}|x\rangle}
where $x$ denotes the totality of scalar fields, $g(\theta )$ is a gauge 
transformation, $\Pi (g)=e^{i\theta C_F }$, and the volume of $U(1)$ is
normalized to unity.  Then one obtains
\eqn\ifor{I=\lim_{\beta\rightarrow 0}\int_{U(1)}d\theta \int dx \langle 
gx|Tr \Pi (g)(-1)^F e^{-\beta H}|x\rangle +\Delta I .}
It has been shown \ms\stern\ that the correction or boundary term of the index
takes the form
\eqn\bound{\Delta I=\lim_{R\rightarrow\infty}\int_{|\vec{x} |=R}
dx\langle gx|
{x^i\over R}\int_{U(1)}Tr\Psi^i (-1)^F Q^{\dagger} H^{-1}\Pi (g)|x\rangle}  
where $H^{-1}$, the inverse of the Hamiltonian, is defined to be zero on
the kernel of $H$, $\Psi^i$ is the fermion coefficient of the 
derivative term in the supercharge $Q$, and $\vec{x}$ is the flat direction
with boundary $|\vec{x}|=R$.  We will not attempt to rigorously 
prove
that this term vanishes for the cases considered here but instead 
will argue that it provides a small correction needed to ensure that
the index is integral.

\subsec{The Calculation}

The calculation of the index for the $n=2$ case is identical to that 
presented in \ms , and we will not belabor the details.  There it was 
established that the index of supersymmetric ground states is one for the 
one-dimensional $U(1)$ gauge theory.  There are also some zero energy modes
decoupled from the gauge theory.  These include modes associated with
the two directions on the intersection of fourbranes and the zero mode
$S^1$ discussed in section three.  We obtain a total of four fermionic 
states and four bosonic states for each supersymmetric ground state of the 
gauge theory.  We need to make sure that the degeneracy of the $S^1$ at the 
intersection does not mess up the counting.  By cutting off the lower
bound on the $y$ integration at $\epsilon$, we can see that there is a
vanishing contribution to the principal index ($I(0)$) from the intersection.
(There are no inverse powers of $|Q_{\pm}|^2$ in the integral over the
charged scalars in the 
correction of order $\epsilon$ to this cutoff.)  We have also taken $y$
to be noncompact to simplify the index calculation.

There are a couple of new details in the $n=3$ calculation.  Fermions from the
constraint $C_F$ are necessary to saturate the fermion zero modes in 
${(-1)}^F$.  We consider $C_F$ as another component of the 
$\sigma\cdot x$ term in $H_F$ in the exponent.  The justification is 
that the commutator terms from rearrangements are higher order in $\beta$
and vanish in the $\beta\rightarrow 0$ limit as discussed in \yi\stern .
The integrand in the
$x^i$ and $\theta$ integrations is then a function of $r'^2=x^i x^i +
\theta^2$.  Let us present the details of this computation.  We start with
\eqn\start{\lim_{\beta\rightarrow 0}\int dxdQ\int_{-\pi}^{\pi}{d\theta
\over 2\pi}{1\over (2\pi\beta)^{d+3/2}}e^{-\sum_m|e^{iq_m\theta}Q_m-Q_m|^2/
\beta}
e^{-\beta V} Tr((-1)^F e^{i\theta C_F} e^{-\beta H_F})}
where $V$ is the bosonic potential, $q_m$ is the charge of a complex
scalar $Q_m$, and the trace is over fermions.
Rescaling all scalars by $\beta^{-1/4}$ yields 
\eqn\resc{\lim_{\beta\rightarrow 0}\int dxdQ\int_{-\pi}^{\pi}{d\theta
\over 2\pi}{\beta^{-d/2-3/4}\over (2\pi\beta)^{d+3/2}}
e^{-\sum_m|e^{iq_m\theta}Q_m-Q_m|^2/
\beta^{3/2}}
e^{- V} Tr((-1)^F e^{i\theta C_F} e^{-\beta^{3/4} H_F}).}
Expanding the $\sum_m|e^{iq_m\theta}Q_m-Q_m|^2/
\beta^{3/2}$ and taking into account the smallness of $\beta$ gives
\eqn\expa{\lim_{\beta\rightarrow 0}\int dxdQ\int_{-\pi}^{\pi}{d\theta
\over 2\pi}{\beta^{-d/2-3/4}\over (2\pi\beta)^{d+3/2}}
e^{-\theta^2\sum_m q_m^2|Q_m|^2/
\beta^{3/2}}
e^{- V} Tr((-1)^F e^{i\theta C_F} e^{-\beta^{3/4} H_F}).}
We might worry that higher order terms in $\theta$ should be included
for the $q_8=-2$ term.  We have explicitly verified by a change of
variables for $\theta$ to extract the contribution near $\theta=\pm\pi$
that this contribution vanishes.
We then rescale $\theta$ by $\beta^{3/4}$ and combine $C_F$ with
$H_F$ to get 
\eqn\comb{\lim_{\beta\rightarrow 0}\int dxdQ\int_{-\infty}^{\infty}{d\theta
\over 2\pi}{\beta^{-d/2}\over (2\pi\beta)^{d+3/2}}
e^{-\theta^2\sum_m q_m^2|Q_m|^2}
e^{- V} Tr((-1)^F  e^{-\beta^{3/4} (H_F+i\theta C_F)}).}
Now by including the appropriate number of fermions, we find that the
$\beta$ dependence has disappeared,
\eqn\bedi{\int dxdQ\int_{-\infty}^{\infty}{d\theta
\over 2\pi}{e^{-(\theta^2 +r^2)\sum_m q_m^2|Q_m|^2}\over (2\pi)^{d+3/2}}
e^{- {1\over 2}(\sum_m q_m|Q_m|^2)^2}{1\over (2d+2)!} Tr((-1)^F  
(H_F+i\theta C_F)^{2d+2}).}
After a calculation we obtain for the fermionic trace the
result 
\eqn\ftr{4(\vec{x}^2+\theta^2)^{d-1}\prod_m q_m^2 (2d+2)! (\sum_m
q_m|Q_m|^2)^2}
where there is an ambiguity for the overall sign that we are taking to
be positive as discussed in section three.  Combining the $x^i$ and
$\theta$ integrations gives
\eqn\combint{4\prod_m q_m^2\int_0^{\infty}dr'{r'^{2d+1}e^{-r^{'2}\sum_m
q_m^2|Q_m|^2}\over (2\pi)(2\pi)^{d+3/2}}\int_{S^3}d\Omega_3
\int dQ (\sum_m
q_m|Q_m|^2)^2 e^{{-1\over 2}(\sum_m
q_m|Q_m|^2)^2} .}
We get
\eqn\resu{{4\prod_m q_m^2 \Gamma (d+1) \pi^2 \over (2\pi)^{(d+3/2)+1}}
\int dQ {(\sum_m
q_m |Q_m|^2 )^2 \over (\sum_m
q_m^2 |Q_m|^2 )^{d+1}} e^{{-1\over 2}(\sum_m
q_m|Q_m|^2 )^2} .}
The remaining integration is done by substituting $Q_m={1\over \sqrt{2}}(Q_{mr}
+iQ_{mi})$ and rescaling $Q_m$ by ${1\over q_m}$ so that the integration is
over real $Q_{mr}$, $Q_{mi}$, and the $\prod_m q_m^2$ factor
disappears.  The remaining computation is straightforward.

We used the computer program Vegas written by G.P. Lepage.  Our computation
involved $10^5$ integrand evaluations per iteration and $10$ iterations.
We obtained the result for the $n=3$ case,
\eqn\indnthr{I(0)=6.0097\pm .0053 .}
Given that the principal
contribution is very close to $6$, we expect a negligible 
boundary contribution.
There are two flat directions, $Q_m =0$ all $m$ or $x^i=0$ and
$\sum_{m} q_m Q_m Q_m^{\dagger}=0$.  An intuitive argument for ignoring
the ``Coulomb'' boundary term is that the charged multiplets become 
very heavy along this direction leaving a free $U(1)$ theory. Unlike
the case of \yi\stern\ we are not starting from a nonabelian theory so there 
is no left over Weyl invariance, and the boundary term should be \yi\
the negative of the principal term for $U(1)$ which vanishes. Without a
superpotential, we cannot ignore the ``Higgs'' boundary.

We will follow somewhat the analysis presented in \stern\ to determine
the asymptotic Hamiltonian in the flat directions for the ground state 
of the massive modes.  To this level of approximation, we will show that 
the boundary term is appropriately small.  We will not ``prove'' that 
the total
index is $6$.

Let us look at the ``Coulomb'' direction.  By a unitary transformation
of the $M_m$ fermions we can write the first term of $H_F$ as
\eqn\fer{r\sum_m q_m (M_{m1}^{\dagger}M_{m1}-M_{m2}^{\dagger}M_{m2})}
where $r=\sqrt{x^i x^i}$.  We decompose the ground state wave function as
\eqn\hogr{\Psi=\Psi_{HO} \Psi_{Flat} (x^i)|F\rangle}
where 
\eqn\hoc{\Psi_{HO}=\prod_m ({r\over |q_m|\pi})^{1\over 4}
e^{-r\sum_m |Q_m|^2 |q_m|}}
and
\eqn\fer{|F\rangle=\prod_m M_{m\alpha_m}^{\dagger}|0\rangle}
where $\alpha_m ={1\over 2}(3+{q_m\over |q_m|})$.
Thus, the supersymmetric harmonic oscillator part of the Hamiltonian 
vanishes on these states.  Next we add up the other contributions.
It is convenient to use harmonic oscillator operators
\eqn\hoop{\eqalign{&Q_{ma}={1\over\sqrt{2r|q_m|}}
(a_{ma}+a_{ma}^{\dagger})\cr
&\partial_{Q_{ma}}=\sqrt{{r|q_m|\over 2}}(a_{ma}-
a_{ma}^{\dagger})\cr}}
with $Q_m={1\over\sqrt{2}}(Q_{mr}+iQ_{mi})$ and
$[a_{ma},a_{nb}^{\dagger}]=\delta_{ab}\delta_{mn}$.  
Note that for
$\Psi_{Flat}=1$, $\langle \Psi|\partial_r|\Psi\rangle =0$.
The net result is that
\eqn\hcoul{H_{Flat}=-{1\over 2}\Delta_x+{d(d+2)\over 8r^2}-{d_+ d_-\over 2r^2}}
where $d$ is the number of chiral multiplets, $d_+$ the number of
positively charged ones, and $d_-$ the number of negatively charged
ones.  We follow \stern\ in realizing that we can lower the ground 
state energy by a redefinition
\eqn\ferp{|F'\rangle=(1-{1\over\sqrt{2} r}\sum_m{q_m\over |q_m|} Q_m
M_m^{\dagger}uL^{\dagger})|F\rangle .}
We then have 
\eqn\hcoullow{H_{Flat}=-{1\over 2}\Delta_x+{d(d-2)\over 8r^2}-
{d_+ d_-\over 2r^2},}
ignoring terms of lower order in $1/r$.
For our case $d_+ =6$, $d_- =2$, $H_{Flat}=-{1\over 2}\Delta_x$, and the $U(1)$
argument \yi\ for a vanishing correction is a good one.

To analyze the ``Higgs'' boundary we first choose a gauge in which $Q_1$
is real.  Then we make the following change of coordinates:
\eqn\change{\eqalign{& Q_1,Q_m\longrightarrow x^0 , Q'_m\cr
&x^0= {\sum_m q_m |Q_m|^2\over
\sqrt{2\sum_m q_m^2 |Q_m|^2}}\cr
&Q'_m= Q_m \, , \quad m>1\cr}.}
The ``Higgs'' branch corresponds to $x^i =x^0 =0$, and the boundary
corresponds to $\nu\rightarrow\infty$ with 
$\nu =\sqrt{2\sum_m q_m^2 |Q_m|^2}$.
Under the change of variables
\eqn\cvar{\eqalign{\sum_m p_m p_m^{\dagger}&=({1\over 2} -{2x^0\over\nu^3}
\sum_m q_m^3 |Q_m|^2 +{{x^0}^2\over \nu^4}\sum_m q_m^4 |Q_m|^2)p^0 p^0\cr
& -i({\sum_m q_m\over\nu}-{2\over\nu^3}
\sum_m q_m^3 |Q_m|^2 -{x^0\over\nu^2}\sum_m q_m^2 +{3x^0\over 
\nu^4}\sum_m q_m^4 |Q_m|^2)p^0\cr
&+2[\sum_m ({q_m\over\nu}-{q_m^2 x^0\over\nu^2})(Q'_mp'_m +
{Q'_m}^{\dagger}{p'_m}^{\dagger})]p^0 +\sum_m p'_m{p'_m}^{\dagger}\cr}.}
Note that $\sum_m q_m=0$ here.  We have not converted all the sums to
primed variables.  To lowest order in $x^0$ and $1/\nu$, the bosonic
Hamiltonian in the massive directions becomes
\eqn\hooph{H_{HO}={1\over 2}(p^i p^i +p^0 p^0)+{1\over 2}\nu^2(Q'_m)
(x^i x^i+x^0 x^0).}
Again we decompose the ground state as
\eqn\grh{\Psi=\Psi_{HO}\Psi_{Flat}(Q'_m)|F\rangle}
where
\eqn\hoh{\Psi_{HO}={\nu (Q'_m)\over\pi}e^{-{\nu x^{\mu}x^{\mu}\over 2}}}
and
\eqn\ferh{|F\rangle={1\over 2}[1-{\sqrt{2}\over\nu}\sum_m q_m Q'_m
M_m^{\dagger}uL^{\dagger}+{1\over\nu^2}{(\sum_m q_m Q'_m
M_m^{\dagger}uL^{\dagger})}^2]|0\rangle .}
Unprimed sums are converted to primed sums by the substitution
\eqn\sub{Q_1^2 =-{1\over q_1}\sum_m q_m {|Q'_m|}^2}
to zeroth order in $x^0$.  By supersymmetry $M_1$ is no longer an
independent fermion on the ``Higgs'' branch.
We find that $(H_{HO}+H_F^2)\Psi=0$ where $H_F^2$ is the term of $H_F$
depending on the $Q'_m$ ($x^0=0$).  Also, the $\sigma\cdot x$ term of
$H_F$ gives zero contribution ($\langle\Psi|H_F^1|\Psi\rangle=0$).
There are many other contributions of order ${1\over \nu^2}$ from
the Hamiltonian.  In addition to the terms in the change of variables,
the Hamiltonian contains some other correction terms, and we find
\eqn\corr{\eqalign{\Delta H  &=q_1^2 x^0 x^0 x^{\mu}x^{\mu}-\sqrt{2}(
{1\over 2}{{x^0}^2\over Q_1}-{1\over 8}{\nu^2{x^0}^2\over q_1^2 Q_1^3})
q_1(M_1^{\dagger}uL^{\dagger}-M_1uL)\cr
&+i({x^0\over\nu^2}\sum_m q_m^2 +{3x^0\over\nu^4}\sum_m (q_m^4-q_1^3
q_m){{|Q'_m|}^2})p^0\cr & -{2x^0\over\nu^2}\sum_m q_m^2 (Q'_m p'_m p^0 +
{Q'_m}^{\dagger} {p'_m}^{\dagger} p^0)+\sum_m p'_m {p'_m}^{\dagger}\cr}}
where we have ignored vanishing or lower order terms.  After some
calculation we have obtained the following result for $H_{Flat}$,
\eqn\corrh{\eqalign{H_{Flat}&=-{1\over 2}\Delta_{Q'_m}+{1\over\nu^4}[
{1\over 162}(|Q'_2|^2+|Q'_3|^2)\cr &+{17\over 18}(|Q'_4|^2+|Q'_5|^2
+|Q'_6|^2)+{67\over 18}|Q'_7|^2+{55\over 6}|Q'_8|^2]\cr}}
where $q_2=q_3=1/3$, $q_4=q_5=q_6=2/3$, $q_7=-1$, and $q_8=-2$.  On the
other hand,
\eqn\ome{\nu^2={4\over 9}(|Q'_4|^2+|Q'_5|^2
+|Q'_6|^2)+{8\over 3}|Q'_7|^2+{28\over 3}|Q'_8|^2 .}
The constraint that $Q_1^2\ge 0$ implies that no $Q'_m$ can have
order greater than $\nu$.  Thus, the term in brackets has order
${1\over \nu^2}$.  We note that the potential $V(Q')>{7\times 10^{-5}\over
r^2}$ where $r=\sqrt{\sum_m |Q'_m|^2}$.

The calculation of the boundary term is complicated by this correction as
well as the constraint that $Q_1^2\ge 0$.  Taking $H_{Flat}$ to be the free
Laplacian, we apply again the argument of \yi ,\foot{I am grateful to A.
Konechny for reminding me of this argument.}
\eqn\delti{\eqalign{\Delta I&=-\lim_{\beta\rightarrow 0}\int dQ'
\int_{-\pi}^{\pi}{d\theta
\over 2\pi}{1\over (2\pi\beta)^7}e^{-\sum_m|e^{iq_m\theta}Q'_m-Q'_m|^2/
\beta}
 Tr((-1)^F e^{i\theta(\sum_m q_m M^{\dagger}_m M_m) } )\cr
&=\pm 2\int dQ'\int_0^{\pi}{d\theta
\over (2\pi)^8}e^{-\sum_m|e^{iq_m\theta}Q'_m-Q'_m|^2}
\prod_m sin^2 (\theta q_m)\cr
&=\pm{1\over 2^7 \pi}\int_0^{\pi}d\theta\prod_m{sin^2 (\theta q_m)\over
(1-cos(\theta q_m))}\cr
&=\pm{1\over\pi}\int_0^{\pi}d\theta cos^2 (\theta)cos^2 ({\theta\over 2})
cos^6 ({\theta\over 3})cos^4 ({\theta\over 6})\cr }.}
The correction computed this way turns out to be
\eqn\corrhiggs{\Delta I=\pm 0.2104 .}
The potential is positive definite and can only decrease the boundary
contribution given that there are no tachyons in this theory.  Also,
the constraint that $Q_1^2\ge 0$ significantly decreases the boundary
contribution.
Given that the total index is integral and our approximations, we have
provided strong evidence that
\eqn\result{I_{n=3}=I(0)+\Delta I=6 .}
We also note that no other integer is consistent with these results.
There is a ${\bf Z}_6$ singularity along $Q_8'=r$, and the $U(1)=S^1$
becomes an $RP^1$ there.  This situation makes the direct boundary
calculation more difficult.
This singularity does not contribute to
the $\beta\rightarrow 0$ calculations.  Had we chosen coordinates
with $Q_8\rightarrow x^0$, we might have avoided this problem.  We are 
somewhat baffled by the role of this singularity in the ``Higgs'' branch.
We have recalculated the boundary term by interchanging $Q_1$ and $Q_8$,
neglecting the asymptotic potential.  We need to compute
\eqn\delih{\eqalign{\Delta I&=\pm{1\over Volume(S^{13})2\pi\cdot 2^{13}}   
\lim_{R\rightarrow\infty}\int_{-\pi}^{\pi}
d\theta\int_{r=R, Q_8^2\ge 0}\prod_m dQ'_m {1\over\sqrt{\sum_m |Q'_m|^2}}\cr
&{Tr((-1)^F e^{i\theta(\sum_m q_m M^{\dagger}_m M_m)})\over (\sum_m |Q'_m|^2
(1-cos(\theta q_m)))^6}\cr}.}
Using Vegas we obtain the result
\eqn\corrhiggsagain{\Delta I=\pm 0.03270\pm .00004.}  
As expected this result is significantly lower than \corrhiggs , and
we expect a further decrease by properly determining the propagator.

A possible heuristic argument for the ``Higgs'' boundary term being negligible
is as follows.  
Suppose we introduce 
a superpotential with infinitesimal gauge invariant cubic and quartic 
couplings.  Including the $D$ term there are enough constraints on the
eight chiral multiplets to lift all of the flat directions so that the
``Higgs'' branch is massive and introduces no boundary correction.  Does
this superpotential change the principal index?  Since these couplings
only multiply the superpotential terms, we cannot do a universal 
rescaling to make them large.  These couplings have the dimension
of mass to some power and are negligible in the high temperature ($\beta
\rightarrow 0$) limit.  Also, the limit that the couplings vanish does
not produce a singularity in the principal index calculation.  In 
conclusion, we argue that the boundary correction from the ``Higgs'' branch
produces a very small correction in the principal index so that the 
index is integral.

\subsec{Counting BPS States}

In the last section we have calculated the index of supersymmetric ground
states in the simplest examples of the $n=2$ and $n=3$ theories.  We will
now use this result along with some plausible assumptions to count the 
BPS ground states and determine the entropy.  Our first assumption is
that the index actually counts the ground states in these theories.  In
any case it counts states that will remain massless under smooth deformations
of the theory.  The degeneracy of states will be bounded from below
by the degeneracies determined from the index.  We have seen in section
two that the Reissner-Nordstrom metric is asymptotically flat.  At large
distances from the intersections, the D-particles experience flat 
ten-dimensional spacetime.  We therefore assume that there is a unique
bound state of $N$ D-particles for every $N$ (as has apparently been
shown as an index \ko ).  Our final assumption is that the D-particles 
and their bound states can interact with any of the $N_1 N_5$
intersections for $n=2$ or $N_2 N_3 N_4$ intersections for $n=3$ to
form the same number of bound states that we have obtained in the one
intersection case.

The $n=2$ index calculation indicates that there are
four massless bosonic modes and
four massless fermionic modes for a D-particle interacting with one 
intersection.  With the above assumptions we write down the following
generating function for the degeneracy of $N_0$ D-particles
interacting with $N_1 N_5$ intersections,
\eqn\zfive{Z=\prod_n{(1+q^n)^{4N_1 N_5}\over (1-q^n)^{4N_1 N_5}}=\sum_{N_0}
d(N_0) q^{N_0}.}
In the above product $n$ indexes the number of D-particles that are bound
together.  The maximum $n$ is $N_0$.  For large $N_0$ this formula implies
$d(N_0)\sim exp(2\pi\sqrt{N_0 N_1 N_5})$,  
exactly the result obtained from the
onebrane-fivebrane system in previous calculations \four\fifteen\sixteen .

The $n=3$ calculation reveals six massless bosonic modes for a D-particle
interacting with one intersection.  Using our assumptions we determine
the generating formula for the degeneracy of $N_1$ D-particles
interacting with $N_2 N_3 N_4$ intersections to be
\eqn\zfour{Z=\prod_n (1-q^n)^{-6N_2 N_3 N_4}=\sum_{N_1}d(N_1) q^{N_1}.}
The maximum $n$ is $N_1$.
Again, we have the previously determined result that $d(N_1)\sim
exp(2\pi\sqrt{N_1 N_2 N_3 N_4})$ \nine\ten\eleven .  Now
that we have a little confidence in our theories, we will see in the next
section what they imply for the quantum mechanics of four-dimensional 
black holes.

\newsec{The Quantum Mechanical System}

\subsec{Generalities}

Let us first write down Lagrangians for the $n=2$ and $n=3$ theories
following from dimensional reduction of four-dimensional theories \wb .
Note again that the $n=3$ theories are the dimensional reduction of anomalous
four-dimensional theories.

For the $n=2$ case we will again assume the superpotential is that given
by $N=2$ supersymmetry in four dimensions (although this is not essential).
We denote neutral scalars in the adjoint of $U(N_0)$ by $Z_{\mu}=
Z_{\mu}^a T^a$ and charged scalars by $A_{\alpha\beta\gamma}$,
$\bar{A}^{\alpha\beta\gamma}$, 
$B^{\alpha\beta\gamma}$, and $\bar{B}_{\alpha\beta\gamma}$
where the first (gauge) index runs from $1$ to $N_0$, the second has
$N_1$ values, and the third $N_5$ values.  We single out the three
components of the four-dimensional gauge field in the dimensional reduction
as $x_i$.  The superpartners of the charged scalars are $\psi_A$ and
$\psi_B$ while those of the neutral scalars are $\lambda_L$ and
$\lambda_N$.  Our hermitian generators $(T^a)_{\alpha}\, ^{\beta}$ of
$U(N_0)$ satisfy $[T^a,T^b]=if^{abc}T^c$ and 
$tr(T^a T^b)={1\over 2}\delta^{ab}$  where $f^{abc}$ are the structure
constants of $SU(N_0)$ (The $U(1)$ generator is 
$(T^{N_0})_{\alpha}\, ^{\beta}={1\over
\sqrt{2 N_0}}\delta_{\alpha}\, ^{\beta}$.).

The Lagrangian is written as follows in the $A_0=0$ gauge.
\eqn\lagtwo{\eqalign{{\cal L}_{n=2}&={1\over 2}\dot{Z}_{\mu}^a\dot{Z}_{\mu}^a
+\dot{\bar A}\dot{A}+\dot{\bar B}\dot{B}
-g^2 tr([Z_{\mu},Z_{\nu}]^2 ) \cr
&-g^2 (|Z_{\mu}A|^2+|Z_{\mu}\bar{B}|^2)-{g^2\over 2}
(\bar{A}T^a A-BT^a\bar{B})^2-2g^2|BT^a A|^2\cr &+i\bar\lambda_L^a\dot
\lambda_L^a+i\bar\lambda_N^a\dot\lambda_N^a+i\bar\psi_A\dot\psi_A
+i\bar\psi_B\dot\psi_B\cr &-g\bar\lambda_L^a[\sigma\cdot x,\lambda_L]^a
-g\bar\lambda_N^a[\sigma\cdot x,\lambda_N]^a
-g\bar\psi_A\sigma\cdot x\psi_A+g\bar\psi_B\sigma\cdot x^T\psi_B\cr &
-\sqrt{2}g(\psi_B uy\psi_A-\bar\psi_A u\bar{y}\bar\psi_B)\cr &
+\sqrt{2}g(\bar{A}T^a\psi_A u\lambda_L^a-\bar\lambda_L^a u\bar\psi_A T^a A
+BT^a\bar\psi_B u\bar\lambda_L^a -\lambda_L^a u\psi_B T^a \bar{B})\cr
&+\sqrt{2} g(\bar{A}T^a\bar\lambda_N^a u\bar\psi_B -\psi_B u\lambda_N^a T^a
A-BT^a\lambda_N^a u\psi_A+\bar\psi_A u\bar\lambda_N^a T^a\bar{B})\cr}}
where $y$, $\bar{y}$ are the components of $Z_{\mu}$ transverse to $x_i$,
and we have suppressed most of the indices.  The $a$ and 
$\mu$ indices should be summed over.  The Gauss' law constraints are
\eqn\contwo{\eqalign{G_{n=2}^a&=-2ig[Z_{\mu},\dot{Z}_{\mu}]^a +ig{d\over dt}
(\bar{A}
T^a A-BT^a \bar{B})\cr &-2g[\bar{\lambda}_L , \lambda_L]^a
-2g[\bar{\lambda}_N , \lambda_N]^a -g\bar{\psi}_A T^a\psi_A+
g\bar{\psi}_B {T^a}^T\psi_B\cr} .}

The $n=3$ Lagrangian is a little easier to write.  We are not including
superpotentials in our analysis here though they may be
significant at higher energies.  The charged scalars 
are $Q_{m\alpha\beta\gamma}
\, ^{\delta}$, $R_n\, ^{\alpha\beta\gamma\delta}$ and hermitian conjugates
where the
first ($\alpha$) index is a $U(N_1)$ gauge index and the $\beta$, $\gamma$, and
$\delta$ indices are indices for the fundamental representations of
$U(N_2)$, $U(N_3)$, and $U(N_4)$; $m$ runs from $1$ to $6$ where the
upper (dual) index can be any of the last three indices; and $n$ runs from
$7$ to $8$.  The charged superpartners are $\psi_m$ and $\psi_n$.
\eqn\lagthree{\eqalign{{\cal L}_{n=3}&={1\over 2}\sum_a\dot x_i^a\dot x_i^a 
+\sum_m\dot{\bar Q}_m
\dot{Q}_m+\sum_n\dot{\bar R}_n\dot{R}_n -g^2 tr([x_i,x_j]^2)\cr
&-g^2(\sum_{m,a} {q_m^a}^2|x_i^a T^a Q_m|^2+\sum_{n,a} {q_n^a}^2
|R_n x_i^a T^a|^2)
-{g^2\over 2}
(\sum_{m,a} q_m^a \bar{Q}_m T^a Q_m+\sum_{n,a} q_n^a R_nT^a\bar{R}_n)^2\cr
&+i\sum_a\bar\lambda_L^a\dot\lambda_L^a +i\sum_m\bar\psi_m\dot\psi_m +i
\sum_n\bar\psi_n\dot\psi_n\cr & -g\sum_a \bar\lambda_L^a
[\sigma\cdot x,\lambda_L]^a
-g\sum_{m,a} q_m^a\bar\psi_m\sigma\cdot x^a T^a\psi_m -
g\sum_{n,a} q_n^a\bar\psi_n\sigma\cdot x^a {T^a}^T\psi_n\cr &
+\sqrt{2}g[\sum_{m,a} q_m^a(\bar{Q}_m T^a\psi_m u\lambda_L^a -
\bar\lambda_L^a u\bar\psi_m T^a Q_m)-\sum_{n,a} q_n^a(R_n T^a\bar\psi_n u\bar
\lambda_L^a
-\lambda_L^a u \psi_n T^a\bar{R}_n)]\cr}}
where the $q_r^a={q_r\over |q_r|}$ for $a\ne N_1$, $q_r^{N_1}=q_r$
with $r=m$ or $n$, and the
$q_m$, $q_n$ have been previously given in section four.  The 
Gauss' law constraints are
\eqn\conthree{\eqalign{G_{n=3}^a&=-2ig[x_i,\dot{x}_i]^a +ig{d\over dt}
(\sum_m q_m^a\bar{Q}_m T^a Q_m +\sum_n q_n^a R_n T^a\bar{R}_n )\cr
&-2g[\bar{\lambda}_L , \lambda_L]^a -g(\sum_m q_m^a\bar{\psi}_m T^a\psi_m
+\sum_n q_n^a\bar{\psi}_n {T^a}^T\psi_n)\cr} .}
The supersymmetries of this action are
\eqn\superthree{\eqalign{\delta_{\eta}x_i^a&=i\bar{\lambda}^a\sigma_i
\eta -i\bar{\eta}\sigma_i\lambda^a\cr
\delta_{\eta}\lambda^a&={i\over 2}g\sigma_{ij}[x_i ,x_j]^a\eta
-ig(\sum_m q_m^a\bar{Q}_m T^a Q_m +\sum_n q_n^a R_n T^a\bar{R}_n )\eta\cr
\delta_{\eta}Q_m&=-\sqrt{2}i\eta u\psi_m\cr
\delta_{\eta}R_n&=-\sqrt{2}i\eta u\psi_n\cr
\delta_{\eta}\psi_m&=\sum_a -igq_m^a\sqrt{2}\sigma\cdot 
x^a T^au\bar{\eta}Q_m
-\sqrt{2}u\bar{\eta}\dot{Q}_m\cr
\delta_{\eta}\psi_n&=\sum_a -igq_n^a\sqrt{2}\sigma\cdot 
x^a {T^a}^T u\bar{\eta}R_n
-\sqrt{2}u\bar{\eta}\dot{R}_n\cr}}
where one needs to use the equations of motion to cancel terms, and $\eta$
is a two-component complex constant fermion.
We have written these Lagrangians in detail for future reference.  Our
analysis from this point will concentrate on the $n=3$ case with some
relevant comments about the $n=2$ case.

\subsec{Reduction to a Conformal Quantum Mechanics}

We will generalize the methods used by \poly\ for reducing a matrix model
to a multidimensional Calogero type model \calo .  We were inspired
in our research by the proposal of \gibto\ that the one-dimensional Calogero
model described the near horizon Reissner-Nordstrom extremal black holes.  
We were unable to confirm their proposal but found evidence from the 
multistrings leading to a generalized Calogero model.  Rewriting the 
bosonic part of
the $n=3$ Lagrangian with some Lagrange multipliers gives the following
result,
\eqn\laglag{\eqalign{{\cal L}^{\Lambda}_{n=3}&={1\over 2}\sum_a\dot{x}_i^a 
\dot{x}_i^a
+\sum_m\dot{\bar{Q}}_m\dot{Q}_m +\sum_n\dot{\bar{R}}_n\dot{R}_n 
+tr(i\Lambda^{(1)}_{ij}
[x_i,x_j]-{1\over 4 g^2}{\Lambda^{(1)}_{ij}}^2\, )\cr &
-\sum_{m,a} {q_m^a}^2(\bar{Q}_m x^a T^a\cdot\Lambda^{(2)}_m+
{\Lambda^{(2)}_m}^{\dagger}\cdot x^a T^aQ_m)
-\sum_{n,a} {q^a_n}^2(R_n x^a T^a\cdot{\Lambda^{(2)}_n}^{\dagger}
+\Lambda^{(2)}_n\cdot x^a T^a\bar{R}_n)
\cr &
+{1\over g^2}(\sum_m|\Lambda^{(2)}_m|^2
+\sum_n|\Lambda^{(2)}_n|^2)\cr &
-\sum_{m,a} q_m^a \bar{Q}_m{\Lambda^{(3)}}^a T^aQ_m-\sum_{n,a} q_n^a R_n
{\Lambda^{(3)}}^a T^a\bar{R}_n
+{1\over 2g^2}\sum_a ({\Lambda^{(3)}}^a)^2\cr}}
where again most of the indices are suppressed.  Integrating out the
$\Lambda$'s gives the bosonic part of the previous Lagrangian \lagthree .

We derive the following equations of motion.
\eqn\motone{\ddot{x}_i^a +i[\Lambda^{(1)}_{ij},x_j]^a +{q_m^a}^2\bar{Q}_m T^a
\Lambda^{(2)}_{mi}+{q_m^a}^2{\Lambda^{(2)}_{mi}}^{\dagger}T^a Q_m +
{q_n^a}^2 R_n T^a
{\Lambda^{(2)}_{ni}}^{\dagger}+{q_n^a}^2\Lambda^{(2)}_{ni}T^a\bar{R}_n=0}
\eqn\motwo{\ddot{Q}_m+\sum_a{q_m^a}^2 x^a T^a\cdot\Lambda^{(2)}_m 
+\sum_a q_m^a
\Lambda^{(3)\,\, a} T^a Q_m=0}
\eqn\mothree{\ddot{R}_n +\sum_a{q_n^a}^2\Lambda^{(2)}_n\cdot x+\sum_a q_n^a 
R_n\Lambda^{(3)\,\, a} T^a=0}
Note that $g^2$ has the dimension of $(mass)^3$.  The low energy limit
corresponds to ignoring the terms of ${\cal L}^{\Lambda}_{n=3}$ with
coupling ${1\over g^2}$.  Doing this, we are left with some constraints.
\eqn\conbosone{[x_i,x_j]=0}
\eqn\conbostwo{\eqalign{\sum_a {q_m^a}^2 x_i^a T^a Q_m&=0\cr 
\sum_a {q_n^a}^2 R_nx_i^a T^a
&=0\cr}}
\eqn\conbosethree{\sum_m q_m^a \bar{Q}_m T^a Q_m +\sum_n q_n^a 
R_n T^a\bar{R}_n=0}
The ``Coulomb'' branch corresponds to setting all the $Q_m$ and $R_n$ to
zero, whereas the ``Higgs'' branch corresponds to setting $x_i$ and the
$D$ constraints \conbosethree\ to zero.

\subsec{The ``Coulomb'' Branch}

The bosonic Lagrangian on the ``Coulomb'' branch is
\eqn\lagc{{\cal L}_{Coul} =tr\lbrace \dot{x}_i^2 +i\Lambda^{(1)}_{ij}[x_i,x_j]
\rbrace .}
We follow \poly\ in deriving a three-dimensional ``spin-Calogero'' \spcalo\
model.  Due to the global $U(N_1)$ symmetry, there is a conserved
matrix,
\eqn\k{V=i\sum_i[x_i,\dot{x}_i].}
Using the constraint \conbosone\ to diagonalize the $x_i$ by a time dependent
unitary matrix $U$, one obtains
\eqn\lagcd{{\cal L}_{Coul}={1\over 2}\sum_{\alpha=1}^{N_1}
{\dot{\vec{q}_{\alpha}}}^2 
+{1\over 2}\sum_{\alpha\ne\beta }{\tilde
{V}_{\alpha\beta}\tilde{V}_{\beta\alpha}\over 
|\vec{q}_{\alpha}-\vec{q}_{\beta}|^2}}
where $\tilde{V}=UVU^{-1}$ and the $\vec{q}_{\alpha}$ are eigenvalues 
of $\vec{x}$.

One also has the relation
\eqn\kvec{\tilde{V}_{\alpha\beta}=i(\vec{q}_{\alpha}-\vec{q}_{\beta})^2
A_{\alpha\beta}}
where $A=\dot{U}U^{-1}$.  This model becomes supersymmetric with the
additional term
\eqn\fercor{{\cal L}^F_{Coul}=2itr(\tilde{\bar{\lambda}} D_t\tilde{\lambda})}
where $\tilde{\lambda}=U\lambda_L U^{-1}$ and $D_t=\partial_t -[A, \,\,]$.
The supersymmetries which leave the action invariant are
\eqn\supcoco{\eqalign{\delta_{\eta}q^i_{\alpha}&=i
\tilde{\bar{\lambda}}_{\alpha\alpha}\sigma_i\eta-i\bar{\eta}\sigma_i
\tilde{\lambda}_{\alpha\alpha}\cr
\delta_{\eta}\tilde{\lambda}&=[(\delta_{\eta}U)U^{-1},\tilde{\lambda}]
\cr}}
where one needs the equations of motion to cancel terms, and the specific
form of $\delta_{\eta}U$ is not required.

The model is invariant under the conformal symmetry 
$SL(2,{\bf R})$ with action
\eqn\confo{\eqalign{&t'={{\bf a}t+{\bf b}\over {\bf c}t+{\bf d}}\cr
&{q_{\alpha}^i}'(t')=q_{\alpha}^i(t)({\bf c}t+{\bf d})^{-1}\cr
&{\bf a}{\bf d}-{\bf b}{\bf c}=1\cr }.}
There is also an $SO(3)$ symmetry (or $SU(2)$ including fermions).  
The conserved angular momentum is $J_{ij}=\sum_{\alpha}(q_{\alpha}^i
\dot{q}_{\alpha}^j-q_{\alpha}^j\dot{q}_{\alpha}^i)$.
The bosonic symmetry of $AdS_2\times S^2$ is $SL(2,{\bf R})\times SO(3)$.  
Since
$g\sim(\alpha')^{-3/4}$ where the string tension is $(2\pi\alpha')^{-1}$,
the near horizon $\alpha'\rightarrow 0$ limit in the supergravity
corresponds to the $g\rightarrow \infty$ limit that we have taken to
derive this theory.  There
is an added result that we can remove one particle far from the others
($|\vec{q}_1|\gg|\vec{q}_{\alpha}|$, $\alpha>1$) and 
obtain a one particle Calogero model,
\eqn\onecal{{\cal L}_1={1\over 2}|\dot{\vec{q}}_1|^2+{L^2+\sum_{\beta\ne 1}
\tilde{V}_{1\beta}\tilde{V}_{\beta 1}\over 2|\vec{q}_1|^2}
+{\cal L}(\vec{q}_{\alpha} ,
\dot{\vec{q}}_{\alpha})_{\alpha>1}}
where $L^2$ is the angular momentum operator for $S^2$.  This result
has previously been obtained by considering a charged particle in the
supergravity background of $AdS_2\times S^2$ \adsone .  Note that the
relativistic corrections found in \adsone\ should result from an $\alpha'$
expansion of the multistring theory. These corrections might
give a clue to stringy excitations of the multistrings.
Their results also indicate that
we should have a nonlinear realization of the supersymmetry.  
 
The bosonic Hamiltonian takes the form
\eqn\hamc{H_{Coul}={1\over 2}\sum_{\alpha=1}^{N_1}
\vec{p}_{\alpha}^{\, 2} 
+{1\over 2}\sum_{\alpha\ne\beta }{\tilde
{V}_{\alpha\beta}\tilde{V}_{\beta\alpha}\over 
|\vec{q}_{\alpha}-\vec{q}_{\beta}|^2}.}
One can write the generators of $SL(2,{\bf R})$ as
\eqn\genes{\eqalign{H&=H_{Coul}\cr D&={-1\over 2}\sum_{\alpha}\vec{p}_{\alpha}
\cdot\vec{q}_{\alpha}\cr K&={1\over 2}\sum_{\alpha}\vec{q}_{\alpha}^{\, 2}\cr}}
satisfying the classical Poisson bracket relations
\eqn\poisslr{\eqalign{\lbrace H,D\rbrace_{PB}&=H\cr
\lbrace K,D\rbrace_{PB}&=-K\cr
\lbrace H,K\rbrace_{PB}&=2D\cr}.}
Classically, we also have the following Poisson bracket relations 
for $\tilde{V}$,
\eqn\poisson{\lbrace \tilde{V}_{\alpha\beta},\tilde{V}_{\gamma\delta}\rbrace
_{PB}=
{-i\over 2}(\delta_{\alpha\delta}\tilde{V}_{\gamma\beta}-\delta_{\beta\gamma}
\tilde{V}_{\alpha\delta}).}
Let $\tilde{V}=-\sum_{a=1}^{N_1^2 -1}\tilde{V}^a T^a$.
Then
\eqn\poisstwo{\lbrace\tilde{V}^a ,\tilde{V}^b\rbrace_{PB}=f^{abc}\tilde{V}^c .}

Let us try to supersymmetrize this Hamiltonian.  There is a problem 
here in that the variation of the constraint under the original
supersymmetries of the ``Coulomb'' branch is nonzero when the 
constraint is applied.
\eqn\suofco{\sigma_i\sigma_j\delta[x_i, x_j]\sim [\lambda,\sigma\cdot x].}
Supersymmetry requires that 
\eqn\suofcotwo{\sigma\cdot(q_{\alpha}-q_{\beta})\tilde
{\lambda}_{\alpha\beta}=0}
or that $\tilde{\lambda}_{\alpha\beta}=0$ for $\alpha\ne\beta$.
Examining the original Lagrangian \lagthree\ with
$Q_m=R_n=0$, we see that we can satisfy the constraint \conbosone\ 
with nonzero
fermions in the conformal limit by imposing another constraint on the
fermions
\eqn\conanot{[\tilde{\bar{\lambda}}^s ,\tilde{\lambda}^t ]_{\alpha\alpha}=0}
all $\alpha$ where $s$, $t$ are $SU(2)$ spinor indices.
The Gauss' law constraint \conthree\ implies
$\tilde{V}=-[\tilde{\bar{\lambda}} ,\tilde{\lambda}]$.  
The internal spin symmetry
will be determined by the fermions, and one will obtain different models
depending on the representation.  Supersymmetry thus implies that the
interaction $\tilde{V}=0$.  We have another option.  By taking a
linear combination of the supersymmetries that breaks the $SU(2)$
symmetry, we can obtain a zero eigenvalue of the analog of \suofcotwo\
and partially preserve the supersymmetry with nonzero interaction.  
This is not an
option if we are to describe a conformal dual theory to quantum gravity 
on $AdS_2\times S^2$.   There may be a way to realize a nontrivial
supersymmetry nonlinearly which is not clear at the moment.

One can rewrite the $\tilde{V}$'s as $SU(q)$ quantum spin degrees of 
freedom \polym .
One sets $\tilde{V}^a=\sum_{m=1}^q \psi_{m\alpha}^{\dagger}T^a_{\alpha\beta}
\psi_{m\beta}$ where $\lbrace\psi_{m\alpha},\psi_{n\beta}^{\dagger}\rbrace=
\delta_{mn}\delta_{\alpha\beta}$.  By defining
\eqn\spin{S^{\alpha}_{mn}=\psi_{m\alpha}^{\dagger}\psi_{n\alpha}-{1\over q}
(\sum_{s=1}^q \psi_{s\alpha}^{\dagger}\psi_{s\alpha})\delta_{mn}}
and using the constraint $\tilde{V}_{\alpha\alpha}=0$ to set 
$\sum_{s=1}^q \psi_{s\alpha}^{\dagger}\psi_{s\alpha}=l$ with $l$ an
integer, the Hamiltonian becomes
\eqn\hamtwo{H_{Coul}={1\over 2}\sum_{\alpha=1}^{N_1}\vec{p}_{\alpha}^{\, 2}-
{1\over 2}\sum_{\alpha\ne\beta}{2 tr(S^{\alpha}S^{\beta})+l(l-q)/q\over
4|\vec{q}_{\alpha}-\vec{q}_{\beta}|^2}.}
The spins are in the $l$-fold antisymmetric representation of $SU(q)$.
One can also obtain an antiferromagnetic interaction by using 
bosonic oscillators.  

The $n=2$ case differs from the $n=3$ case by extra global $U(1)^2$
symmetry.  This symmetry originates from the extra BPS
deformation directions for the $n=2$ multistrings.

\subsec{The ``Higgs'' Branch}

The bosonic Lagrangian on the ``Higgs'' branch is 
\eqn\lagh{{\cal L}_{Higgs}=\dot{\bar{Q}_m}\dot{Q}_m+\dot{\bar{R}_n}\dot{R}_n
-\sum_{m,a} q_m^a\bar{Q}_m{\Lambda^{(3)}}^a T^a Q_m - 
\sum_{n,a} q_n^a R_n{\Lambda^{(3)}}^a T^a\bar{R}_n}
where the $D$ constraints are enforced by $\Lambda^{(3)}$.  The Gauss' law 
constraints are 
\eqn\gaussh{i{d\over dt}
(\sum_{m,a} q_m^a\bar{Q}_m T^a Q_m +\sum_{n,a} q_n^a R_n T^a\bar{R}_n )
=(\sum_{m,a} q_m^a\bar{\psi}_m T^a\psi_m
+\sum_{n,a} q_n^a\bar{\psi}_n {T^a}^T\psi_n).}
If $\Lambda^{(3)}$
were time independent, this system would be a simple harmonic oscillator.  
One can see 
that $\Lambda^{(3)}$ has the dimension of $(mass)^2$.  For this
action to be conformally invariant, we require that $\Lambda^{(3)}$
transforms as
\eqn\trla{{\Lambda^{(3)}}'=(\gamma t +\delta)^4\Lambda^{(3)}.}
If we redefine the chiral multiplets by a time dependent unitary matrix
that diagonalizes $\Lambda^{(3)}$ and introduces a covariant time
derivative, we have an interpretation of the 
square roots of the $N_1$
eigenvalues as time dependent inverse scale sizes of the $N_1$ 
D-particles in the transverse 
dimensions.  To supersymmetrize the ``Higgs'' branch, one can perform a
change of coordinates that is similar to \change\ (assuming
$8N_1 N_2 N_3 N_4>N_1^2$) so that the complex dimension is 
$8N_1 N_2 N_3 N_4-N_1^2$.
The remaining massless
modes have superpartners with the Lagrangian 
\eqn\ferhi{{\cal L}_{Higgs}^F =i\sum_m\bar{\psi}_m\dot{\psi_m}+
i\sum_n\bar{\psi}_n\dot{\psi_n}.}
The action is invariant under the supersymmetries
\eqn\suph{\eqalign{\delta_{\eta}Q_m&=-\sqrt{2}i\eta u\psi_m\cr
\delta_{\eta}R_n&=-\sqrt{2}i\eta u\psi_n\cr
\delta_{\eta}\psi_m&=-\sqrt{2}u\bar{\eta}\dot{Q}_m\cr
\delta_{\eta}\psi_n&=-\sqrt{2}u\bar{\eta}\dot{R}_n\cr}.}
In the limit in which the entropy estimate of section 4.3 is valid, 
$N_1\gg N_2 N_3 N_4$ so the ``Higgs'' branch is massive.
In the $g\rightarrow\infty$ limit the ``Coulomb'' and
``Higgs'' branches appear to be decoupled from each other.  At higher 
energies the two branches are coupled through the harmonic oscillator
modes that have been ignored in the conformal limit.

\newsec{Discussion}

We have conjectured that D-particles at D-brane intersections form
multistrings and that these multistrings are the relevant degrees of freedom
of black holes formed from these intersections.  An index calculation 
shows that the counting of states is correct for one D-particle 
interacting with one intersection.  With several assumptions, one sees
that these multistrings can account for the ground state entropy of the
black hole.  We have derived from the multistring theory a conformal
quantum mechanics, the ``Coulomb'' branch, that exhibits some of the expected
properties of supergravity on $AdS_2\times S^2$.  We have also derived
another conformal quantum mechanics, the ``Higgs'' branch, that describes
part of the moduli space of D-particle--D-intersections.  
These two theories are
coupled at higher energies.  We expect also at higher energies stringy 
corrections to the low energy multistring theory will play a role though we
currently don't know how to describe these excitations.  A future goal is
to determine the full effective theory of the multistrings.

It would be interesting to see whether one could reproduce the BPS
spectrum of supergravity on $AdS_2\times S^2$ \adsbps\ from the conformal
mechanics.\foot{There is also related work \adsact\ on the construction of a 
superstring action on
$AdS_2\times S^2$.}  The bosonic symmetries are the same, and it remains to
see how and whether supersymmetry can be realized.  It would also
be interesting to compare correlation functions in the two theories.
An even more interesting and current project is to see whether the full quantum
mechanics describes the dynamics of the nonextremal Reissner-Nordstrom
black hole at low energies where stringy corrections can be neglected.

\bigskip\centerline{\bf Acknowledgments}\nobreak

I would like to thank P. Aschieri, K. Bardakci, C. Csaki, R. Dawid, 
S. Giddings, S. Gopalakrishna, M. Halpern, K. Hori,
S. Kachru, C. Kolda, A. Konechny, B. Morariu, H. Ooguri, J. Terning,
and B. Zumino for sharing some of their 
time to listen to and
comment helpfully on some of my ideas. I am also grateful to J. Anderson,
A. Friedland, S. Gopalakrishna, I. Hinchliffe, and C. Kolda
for important help with various computer programs.  This research 
was supported in part 
by the Director, 
Office of Science,
Office of Basic Energy Services, of the U.S. Department of Energy under
contract DE-AC03-76SF00098.

\listrefs
\end